\documentclass[aps,11pt,tightenlines,nofootinbib,prd,superscriptaddress]{revtex4-2}
\usepackage{natbib}

\usepackage{lipsum}
\usepackage{xcolor}
\usepackage{graphicx}
\usepackage{caption}
\usepackage{subcaption}
\usepackage{dcolumn}
\usepackage{bm}
\usepackage{amssymb}
\usepackage{amsmath}
\usepackage{mathrsfs}
\usepackage{physics}
\usepackage{graphicx,color} 
\usepackage[english]{babel}     
\usepackage{graphicx,wrapfig,lipsum}      
\usepackage{multimedia}    
\usepackage[mathscr]{eucal}  
\usepackage{bm}              
\usepackage{amssymb}           
\usepackage{amsmath}
\makeatletter
\newcommand*{\encircled}[1]{\relax\ifmmode\mathpalette\@encircled@math{#1}\else\@encircled{#1}\fi}
\newcommand*{\@encircled@math}[2]{\@encircled{$\m@th#1#2$}}
\newcommand*{\@encircled}[1]{%
  \tikz[baseline,anchor=base]{\node[draw,circle,outer sep=0pt,inner sep=.2ex] {#1};}}
\usepackage{tikz}
\usepackage{tikz-3dplot}
\usetikzlibrary{decorations.markings}
\usepackage[outline]{contour} 
\tikzset{>=latex} 
\makeatother

\begin{document}

\title{Geometric R\'enyi mutual information induced by localized particle excitations in quantum field theory}   

\author{Willy A. Izquierdo}
\email{willy.izquierdo@unesp.br}
\affiliation{Instituto de F{\'i}sica Te{\'o}rica, Universidade Estadual Paulista, Rua Dr. Bento Teobaldo Ferraz, 271 - Bloco II, 01140-070 S{\~a}o Paulo, SP, Brazil}

\author{David R. Junior}
\email{davidjunior@id.uff.br}
\affiliation{Instituto de F{\'i}sica Te{\'o}rica, Universidade Estadual Paulista, Rua Dr. Bento Teobaldo Ferraz, 271 - Bloco II, 01140-070 S{\~a}o Paulo, SP, Brazil}
\affiliation{Institut f\"ur Theoretische Physik, Universit\"at T\"ubingen, Auf der Morgenstelle 14, 72076 T\"ubingen, Germany}
\affiliation{
Instituto de F\'\i sica, Universidade Federal Fluminense, 24210-346 Niter\'oi, RJ, Brazil.}
\author{Gast\~ao  Krein}
\email{gastao.krein@unesp.br}
\affiliation{Instituto de F{\'i}sica Te{\'o}rica, Universidade Estadual Paulista, Rua Dr. Bento Teobaldo Ferraz, 271 - Bloco II, 01140-070 S{\~a}o Paulo, SP, Brazil}

\date{\today}

\begin{abstract}     
Quantum field theory exhibits rich spatial correlation structures even in the vacuum, where entanglement entropy between regions scales with the area of their shared boundary. While this vacuum structure has been extensively studied, far less is understood about how localized wave packets influence correlations between field values in different spatial regions. In this work, we use the Schr\"odinger representation to study the R\'enyi mutual information between complementary spatial regions for a localized wave packet of a free massless scalar field in $(d+1)$ dimensions. We find that the mutual information in this excited state includes both a vacuum term and an excitation-induced contribution. To obtain quantitative results, we specialize to $1+1$ dimensions and evaluate the R\'enyi-2 mutual information between the negative and positive halves of the real line. We find that the excitation generates finite, positive correlations that are maximized when the wave packet sits at the boundary and decrease with its distance from it, at a rate determined by the wave packet's width.Our findings constitute a contribution toward a more comprehensive understanding of quantum correlations in multiparticle states with nontrivial spatial localization, analyzed within a quantum field-theoretical framework.
 
\end{abstract}

\maketitle

\section{Introduction}

The study of geometric correlations in quantum field theory (QFT) has its roots in attempts to understand black‐hole entropy and vacuum entanglement. In particular, the spatial structure of entanglement in quantum fields has become a central theme in areas such as holography, quantum information, and condensed matter physics. Quantifying how field degrees of freedom in different regions of space are correlated has offered deep insights into the nature of quantum matter and spacetime. In a seminal work, Srednicki studied the entanglement between the degrees of freedom of a spatial region $\Omega$ and its complement $\overline{\Omega}$ for the ground state (the vacuum state) of a massless scalar field and found that the leading contribution scales with the area of the boundary~$\partial \Omega$~\cite{Srednicki1993}. In the years that followed, many authors (see the reviews \cite{HOLZHEY1994443,PasqualeCalabrese_2004,Casini_2009,Nishioka:2018khk}) generalized these results. It is now understood that for the vacuum state of a free field theory, the entanglement entropy $S(\Omega)$ between the degrees of freedom of a spatial region $\Omega$ and its complement diverges with the ultraviolet (UV) cutoff in the form of an area term plus subleading divergences. That is, one finds in $d$ dimensions 
\begin{equation}
S(\Omega) \sim  \alpha  \frac{|\partial \Omega|}{\epsilon ^{d-1}} +\cdots +a_{\log}\ln\frac{L}{\epsilon } +\cdots ,\label{arealaw}
\end{equation} 
where $\epsilon $ is a short‐distance regulator, and $a_{\log}$ is a universal coefficient. In $1+1$ dimensions, conformal field theory predicts $a_{\log} =c/3$, so that $S=(c/3)\ln (L/\epsilon )+$const \cite{HOLZHEY1994443,PasqualeCalabrese_2004}. In higher dimensions, one similarly finds an area‐law divergence plus logarithmic terms whose coefficients depend on conformal coefficients~\cite{Casini_2009}. Moreover, other entanglement measures, such as the Rényi entropies, exhibit similar behavior \cite{PhysRevD.89.125016}.

Another way to characterize correlations among field variables in complementary spatial regions is through mutual information~\cite{nielsen00,Groisman:2005,Wolf:2007tdq}. In particular, one can use a generalized R\'enyi mutual information of order~$n$, defined as:
\begin{align}
    I^{(n)}(\Omega,\bar{\Omega})= S^{(n)}(\Omega)+S^{(n)}(\bar{\Omega})-S^{(n)}(\Omega\cup\bar{\Omega})\;, \label{def-mut}
\end{align}
where $S^{(n)}(\Omega)$ is the R\'enyi-n entropy associated with the reduced probability distribution $P_\Omega[g_\Omega]$ to measure the field value $g_\Omega$ in $\Omega$, irrespective of its value in $\bar{\Omega}$. This entropy is defined in Eq.~\eqref{S-n} and reduces to the Shannon entropy in the limit $n \rightarrow 1$. The reduced distribution itself is given by
\begin{align}
    P_\Omega[g_\Omega] = \int [D\phi]_{g_\Omega} P[\phi] \makebox[.5in]{,} P[\phi] = |\Psi(\phi)|^2\;,
\end{align}
where \(\Psi(\phi) = \langle \phi|\Psi\rangle\) is the wave functional in the field basis. The mutual information \(I^{(n)}(\Omega, \bar{\Omega})\) thus quantifies the correlation between the random variables $g_\Omega, g_{\bar{\Omega}}$, i.e., the field values in $\Omega$~and~$\bar{\Omega}$. This quantity is closely related to that introduced in~\cite{Scalet2021computablerenyi}, where \(S^{(n)}\) denotes the Rényi entropy and recovers the von Neumann mutual information in the limit $n \rightarrow 1$. Unlike the von Neumann entropy, however, the Rényi mutual information defined in Eq.~\eqref{def-mut} is explicitly basis dependent. Once a basis is fixed, such as the field basis used in geometric entropy studies, it provides a meaningful measure of spatial correlations. As a side note, although field operators are correlated in position space, their momentum modes in a free field theory are entirely uncorrelated~\cite{PhysRevD.86.045014,Hsu:2012gk,Costa:2022bvs}.

We also note that the dynamics of one-dimensional superfluids can be effectively described by a Klein--Gordon field. In particular, consider two adjacent $1D$ Bose gases realized with ultracold atoms. The spatially resolved relative phase between the two condensates, $\varphi(z)=\theta_1(z)-\theta_2(z)$, is well described in the low--energy regime by the sine--Gordon Hamiltonian. In the limit of strong tunneling between the condensates, this Hamiltonian reduces to that of a massive Klein--Gordon field. It was shown in Ref.~\cite{Gluza2020} that the covariance matrix of this system can be reconstructed through a tomography procedure. In Ref.~\cite{Tajik2023}, this method was applied to a system divided into a subsystem $A$ and its complement $\bar{A}$. The authors verified the area law for the basis--independent R\'enyi mutual information at finite temperature. Within the same framework, the basis--dependent R\'enyi mutual information could in principle be obtained more conveniently, since it depends only on the diagonal elements of the density matrix in the field basis. However, accessing the vacuum state experimentally within this setup is challenging. In the experiment reported in Ref.~\cite{Tajik2023}, the finite temperature of the Bose gas ($10$--$100,\mathrm{nK}$) leads to occupation numbers of the low--momentum modes that remain significantly above their ground--state values. 

The R\'enyi mutual information in the position-space field basis was studied in the context of the vacuum state of quantum fields. In particular, the Shannon mutual information for a free scalar field also follows an area law in any dimension~\cite{David}. This quantity has been extensively studied in one-dimensional quantum spin chains in Refs.~\cite{PhysRevB.80.184421,PhysRevB.84.195128,PhysRevLett.107.020402}, where it proved to be a useful tool for the classification of one- and two-dimensional quantum critical points. Moreover, it has been analyzed in a chain of harmonic oscillators in Ref.~\cite{PhysRevB.88.045426}, where it was found to coincide with the \(n=2\) R\'enyi entanglement entropy. Its natural generalization, the R\'enyi mutual information, was investigated in Refs.~\cite{AlcarazRajabpour2013,AlcarazRajabpour2015} in critical spin chains, where it was shown to encode universal information about the underlying conformal field theory (CFT). In contrast to the extensive literature on entanglement entropy, these field correlations for excited states remain comparatively unexplored. Entanglement in excited states generated by primary operators in CFT were studied in Refs. \cite{PhysRevLett.106.201601,palmai:2014}. Furthermore, Refs.~\cite{castroalvaredoI_2018,castroalvaredoII_2019,castroalvaredoIII_2019} investigated entanglement properties in one-particle and multiparticle states with a definite rapidity. More recently, Refs. \cite{Zhang2020, Zhang2021} studied entanglement in multiparticle states with well defined momenta in a free massless scalar theory. Low-energy excited states were also examined within an holographic approach in Ref. \cite{Blanco2013} using a perturbative approach, and a direct calculation of entanglement in these states was carried out in Ref.~\cite{Wong2013}. In the latter works, agreement was found that for a spatially uniform energy density, while a discrepancy between the field-theoretic and the holographic results was identified for the nonuniform case. 

On the other hand, wave packets, which are excited states with a localized probability distribution in position space, obtained by acting with position-space creation operators on the vacuum state, have not been studied in either of these settings. The present work is devoted to filling this gap by investigating how the precise location of a wave packet modifies the mutual information between complementary regions in a free massless scalar field theory. We employ the Schrödinger representation to construct such states by acting with localized creation operators on the vacuum and compute the induced corrections to the R\'enyi mutual information. This framework provides a foundation for analyzing correlations in more general multiparticle states formed by position-space wave packets, which have predominantly been studied within nonrelativistic quantum mechanics, and may offer a better understanding of entanglement dynamics from a field-theoretic perspective.

In section \ref{sec:one-particle} we review one-particle states for the free scalar field in the Schr\"odinger picture. In section \ref{sec:generalization}, the methods developed in \cite{David} to obtain the Shannon mutual information between two complementary regions $\Omega,\bar{\Omega}$ for the vacuum state are reviewed and generalized for the computation of Rényi-n entropies and the corresponding mutual information of one-particle states. In section \ref{sec:explicit} the Rényi-2 mutual information of one-particle states is explicitly computed for the particular one-dimensional case with $\Omega=(-\infty,0]$ and $\bar{\Omega}=[0,\infty)$. Finally, in section \ref{sec:conclusions} we present our conclusions.

\section{Localized one-particle states}
\label{sec:one-particle}
A free real scalar field $\phi (x)$ in $d+1$-dimensions is described by the action 
\begin{equation}
S[\phi ]=\frac{1}{2}\int d^{d+1} x\ \left( \partial _{\mu } \phi \partial ^{\mu } \phi -m^{2} \phi ^{2}\right) .
\end{equation}
Canonical quantization promotes $\phi (x)$ and its conjugate momentum $\pi (x)$ to operators $\hat{\phi }$ and $\hat{\pi }$, satisfying equal-time commutation relations. In the Schrödinger picture, operators are time-independent, and the state is described by a wavefunctional $\Psi [\phi ,t]=\langle \phi |\Psi (t)\rangle $, satisfying the Schrödinger equation
\begin{align}\label{Schrödinger equation}
i\partial _{t} \Psi [\phi ,t]=\hat{H} \Psi [\phi ,t],
\end{align}
where $\hat{H}$ is the Hamiltonian operator of the field theory. In this picture, $\hat{\phi } (x)$ acts as a multiplication operator and $\hat{\pi } (x)$ as a functional derivative:
\begin{equation}
\hat{\phi } (x)\Psi [\phi ]=\phi (x)\Psi [\phi ],\ \ \ \ \ \hat{\pi } (x)\Psi [\phi ]=-i\frac{\delta }{\delta \phi } \Psi [\phi ]\;.
\end{equation}

The ground state wave functional $\Psi _{0} [\phi ]=\bra{\phi }\ket{\Psi _{0}}$ for the free scalar field is given by the Gaussian
\begin{equation}\label{ground state wave functional}
\Psi _{0} [\phi ]=\left[\operatorname{det}\left(\sqrt{-\nabla ^{2}+m^2}\right)\right]^{1/4}\exp\left( -\frac{1}{2}\int d^{d}\mathbf{x} \ \phi (\mathbf{x} )[\mathcal{O} \phi ](\mathbf{x} )\right)\;, 
\end{equation}
 where $\mathcal{O} =\sqrt{-\nabla ^{2} +m^{2}}$ is the fractional Laplacian operator. The fractional Laplacian can be defined in several equivalent ways \cite{Caffarelli,Landkof1972}. An illuminating one is given in terms of an integral kernel, which, for $m=0$, reads
\begin{gather}
    (-\nabla)^{2\gamma}\phi(x)=C_{2\gamma}\int dy \frac{\phi(x)-\phi(y)}{|x-y|^{2\gamma+d}}\makebox[.5in]{,}C_{2\gamma}=\frac{2^{2\gamma-\frac{d}{2}}\Gamma{\left(\frac{2\gamma+d}{2}\right)}}{\pi^{\frac{d}{2}}\Gamma(-\gamma)}\;.
\end{gather}
For $\gamma=1/2$, the operator is strongly nonlocal. Such nonlocality is responsible for inducing correlations between the degrees of freedom associated with spatially separated regions.

The annihilation and creation operators are functional operators defined as (see \cite{Brian_Hatfield})
\begin{equation}
\hat{a} (\mathbf{x} )=\frac{1}{\sqrt{2}}\left(\mathcal{O}^{1/2} \phi (\mathbf{x} )+\mathcal{O}^{-1/2}\frac{\delta }{\delta \phi (\mathbf{x} )}\right) ,\ \ \ \ \ \ \ \hat{a}^{\dagger } (x)=\frac{1}{\sqrt{2}}\left(\mathcal{O}^{1/2} \phi (\mathbf{x} )-\mathcal{O}^{-1/2}\frac{\delta }{\delta \phi (\mathbf{x} )}\right)\;,
\end{equation}
which satisfy the algebra 
\begin{equation}
[\hat{a} (\mathbf{x} ),\hat{a}^{\dagger } (\mathbf{x} ')]=\delta ^{(d)} (\mathbf{x} -\mathbf{x} '),\ \ \ \ \ [\hat{a} (\mathbf{x} ),\hat{a} (\mathbf{x} ')]=[\hat{a}^{\dagger } (\mathbf{x} ),\hat{a}^{\dagger } (\mathbf{x} ')]=0\ .
\end{equation}
The operator $\hat{a}(x)$ annihilates the vacuum state \eqref{ground state wave functional}, while $\hat{a}^\dagger(x)$ creates excited states. A localized one-particle state is constructed applying creation operators weighted by a wave packet $ f( t,\mathbf{x})$: 
\begin{equation}
\Psi _{1} [\phi ]=\langle \phi |\Psi _{1} \rangle =\int d^{d}\mathbf{x} \ f(\mathbf{x} ,t)\ \hat{a}^{\dagger } (\mathbf{x} )
\Psi _{0} [\phi ],\ \ \ \ \ \ \ \int d^{d}\mathbf{x} \ |f(\mathbf{x} ,t)|^{2} =1\ ,
\end{equation}
where the last condition has to be imposed to ensure that the state is normalized. In addition, $f(\mathbf{x} ,t)$ has to satisfy 
\begin{equation}\label{Shrödiger equation to wavepacket}
i\partial _{t} f(\mathbf{x} ,t)=\sqrt{-\nabla ^{2} +m^{2}} f(\mathbf{x} ,t)
\end{equation}
to ensure that the Schrödinger equation \eqref{Schrödinger equation} is satisfied. With these definitions, the following one-particle wavefunctional is obtained 
\begin{equation}
\Psi _{1} [\phi ]=\sqrt{2}\int d^{d}\mathbf{x} \ f(t,\mathbf{x} )\ [O^{1/2} \phi ](\mathbf{x} )\Psi _{0} [\phi ],
\end{equation}
where $O^{1/2}$ is the square-root of the fractional Laplacian $\mathcal{O}$. Thus, the excited-state probability is
\begin{equation}\label{excited-state probability}
P_{1} [\phi ]=| \Psi _{1} [\phi ]| ^{2} =2\Bigl|\int d^{d}\mathbf{x} \ f(t,\mathbf{x} )\ [O^{1/2} \phi ](\mathbf{x} )\Bigl|^{2} P_{0} [\phi ],
\end{equation}
where $P_{0} [\phi ]$ is the vacuum probability functional. This form shows that $P_{1} [\phi ]$ differs from the vacuum $P_{0} [\phi ]$ by a quadratic functional of the field weighted by the packet profile $f(t,\mathbf{x} )$.

\section{R\'enyi-\lowercase{n} entropies for a massless scalar field}
\label{sec:generalization}
Entanglement entropy and R\'enyi mutual information are complementary measures of correlations in quantum field theory (QFT). The entanglement entropy between the degrees of freedom in a region $A$ and its complement is obtained by the von Neumann entropy of the reduced density matrix $\rho _{A} =\text{Tr}_{\overline{A}} |\Psi \rangle \langle \Psi |$,
\begin{equation}
S_{A} =-\text{Tr}_{A} (\rho _{A}\ln \rho _{A} )\;,
\end{equation}
and is explicitly basis-independent. In continuum QFT, this geometric entropy typically satisfies an “area law” (diverging as the boundary area of $A$). In contrast, the R\'enyi entropy measures the information of a probability distribution, which, in our case, has a purely quantum origin, given by $P[\phi ]=|\Psi [\phi ]|^{2}$. We are ultimately interested in computing the R\'enyi-n mutual information between the regions $\Omega, \bar{\Omega}$, given by
\begin{align}
    I^{(n)}(\Omega:\bar{\Omega})=S^{(n)}(\Omega)+S^{(n)}(\bar{\Omega})-S^{(n)}(\Omega\cup \bar{\Omega})\;, \label{mutualinfo}
\end{align}
where $S^{(n)}(\Omega)$ and $S^{(n)}(\bar{\Omega})$ are the R\'enyi-$n$ entropies associated with the reduced probability distributions $P[g_\Omega]$ and $P[g_{\bar{\Omega}}]$ for measuring the values $g_\Omega$ and $g_{\bar{\Omega}}$ of the field in $\Omega$ and $\bar{\Omega}$, irrespective of their values at $\bar{\Omega}$ and $\Omega$, respectively. The quantity \eqref{mutualinfo} measures the correlation between the random variables $g_\Omega, g_{\bar{\Omega}}$, and vanishes in the absence of entanglement between the degrees of freedom in $\Omega, \bar{\Omega}$.

\subsection{Reduced Probabilities and the Reduced Density Matrix}

The reduced probability density $P_{\Omega} [g_{\Omega} ]$ quantifies the likelihood of observing a field configuration $g_{\Omega} (\mathbf{x} )$ in a spatial region $\Omega$, irrespective of the field value in the complementary region $\overline{\Omega}$. \ It is obtained by functionally integrating the probability distribution $P[\phi ]=|\Psi [\phi ]|^{2}$ over $\overline{\Omega}$ while fixing $\phi (\mathbf{x} )=g_{\Omega} (\mathbf{x} )$ in $\Omega$, i.e.
\begin{equation}
P_{\Omega} [g_{\Omega} ]=\int [\mathcal{D} \phi ]_{g_{\Omega}} P [\phi ]\;.
\end{equation}
The measure $[\mathcal{D} \phi ]_{g_{\Omega}}$, which performs the integration over the degrees of freedom in $\bar{\Omega}$ with $g_\Omega$ fixed, satisfies the following composition property: 
\begin{equation}
\int [\mathcal{D} \phi ]=\int [\mathcal{D} g_{\Omega} ]\int [\mathcal{D} \phi ]_{g_{\Omega}} .
\end{equation}

To compute the reduced probability density $P_{0\Omega} [g_{\Omega} ]$ for the vacuum state, we can follow \cite{David} and use the saddle-point method, which gives us that
\begin{equation}\label{reduced probability for the vacuum state}
P_{0\Omega} [g_{\Omega} ]=\left[\frac{\det (\mu ^{-1}\mathcal{O} )}{\det_{\overline{\Omega }} (\mu ^{-1}\mathcal{O} )}\right]^{1/2}\exp\left( -\int d^{d}\mathbf{x} \ \phi _{0\Omega} (\mathbf{x} )[\mathcal{O} \phi _{0\Omega} ](\mathbf{x} )\right) ,
\end{equation}
where $\mathcal{O}=\sqrt{-\nabla^2}$, $\mu $ is a mass scale introduced for dimensional consistency, it contributes in the vacuum correlations and drops out in the particle induced correlations, and $\det_{\overline{\Omega}}$ denotes the determinant restricted to the region $\overline{\Omega}$. Also, the saddle-point field configuration, $\phi _{0\Omega} (\mathbf{x} )$, satisfies the nonlocal Poisson problem
\begin{align}
 [\mathcal{O} \phi _{0\Omega }](\mathbf{x}) = & 0,\ \ \ \ \ \ \ \ \ \ \ \ \mathbf{x} \in \overline{\Omega }\nonumber \\
\phi _{0\Omega }(\mathbf{x}) = & g_{\Omega }(\mathbf{x}) ,\ \ \ \ \ \mathbf{x} \in \Omega .
\end{align}
The solution to this problem is \cite{Landkof1972,doi:10.1142/S021949370500150X,CAFFARELLI2016767}
\begin{equation}\label{homogeneous Poisson problem solution}
\phi _{0\Omega} (\mathbf{x} )=\begin{cases}
\int _{\Omega} d^{d}\mathbf{y} \ \mathcal{P} (\mathbf{x} ,\mathbf{y} )g_{\Omega} (\mathbf{y} ), & \mathbf{x} \in \overline{\Omega}\\
g_{\Omega} (\mathbf{x} ), & \mathbf{x} \in \Omega.
\end{cases}
\end{equation}
where $\mathcal{P} (\mathbf{x} ,\mathbf{y} )$ is the Poisson kernel.

Now, let us study the reduced probability density $P_{1\Omega} [g_{\Omega} ]$ for the excited state. Again, we may use the saddle-point method, and the result is
\begin{equation}\label{reduced probability for the localized excited state}
P_{1}{}_{\Omega} [g_{\Omega} ]=2\left[\frac{\det\left( \mu ^{-1}\mathcal{O}\right)}{\det_{\overline{\Omega}}\left( \mu ^{-1}\mathcal{O}\right)}\right]^{1/2}\left. \left(\frac{\delta ^{2}}{\delta \alpha ^{2}} -\frac{\delta ^{2}}{\delta \beta ^{2}}\right)\exp( W[\phi _{1\Omega} ,f_{\alpha ,\beta } ])\right| _{\alpha =\beta =0} ,
\end{equation}
where $f_{\alpha,\beta}(t,\mathbf{x}) = \alpha  \Re[ f(t,\mathbf{x})] + i\beta  \Im[f(t,\mathbf{x})]$ and
\begin{equation}
W[\phi _{1\Omega} ,f_{\alpha ,\beta } ]\equiv -\int d^{d}\mathbf{x} \ \left( \phi _{1\Omega} (\mathbf{x} )[\mathcal{O} \phi _{1\Omega} ](\mathbf{x} )-f_{\alpha ,\beta } (t,\mathbf{x} )\left[\mathcal{O}^{1/2} \phi _{1\Omega}\right] (\mathbf{x} )\right) ,
\end{equation}
and the saddle-point field $\phi _{1\Omega} (\mathbf{x} )$ satisfies a nonlocal Poisson problem with a source term related to $\left[\mathcal{O}^{1/2} f_{\alpha ,\beta }\right] (t,\mathbf{x} )$ 
\begin{align}
[\mathcal{O} \phi _{1\Omega} ](\mathbf{x} ) & =\frac{1}{2}\mathcal{O}^{1/2} f_{\alpha ,\beta } (t,\mathbf{x} ),\ \ \ \ \ \ \ \ \ \ \ \ \mathbf{x} \in \overline{\Omega} \nonumber \\
\phi _{1\Omega} (\mathbf{x} ) & =g_{\Omega} (\mathbf{x} ),\ \ \ \ \ \ \ \ \ \ \ \ \ \ \ \ \ \ \ \ \ \ \ \ \ \ \ \mathbf{x} \in \Omega
\end{align}
with solution 
\begin{equation}\label{non homogeneous Poisson problem solution}
\phi _{1\Omega} (\mathbf{x} )=\phi _{0\Omega} (\mathbf{x} )+\frac{1}{2}\int _{\overline{\Omega}} d^{d}\mathbf{y} \ G(\mathbf{x},\mathbf{y})\left[\mathcal{O}^{1/2} f_{\alpha ,\beta }\right] (t,\mathbf{y} ),\ \ \ \ \ \ \ \ \ \ \mathbf{x} \in \overline{\Omega} \ 
\end{equation}
where $G(\mathbf{x},\mathbf{y})$ is the Green's function of the problem (see \cite{ClaudiaBucur}).

\subsection{R\'enyi-n entropy}

 For a probability distribution $P_\Omega[g_\Omega]$, the R\'enyi-$\displaystyle n$ entropy $S^{( n)}$ is defined as 
\begin{equation}
S^{( n)}( \Omega ) =\frac{1}{1-n}\ln\int [Dg_{\Omega} ]\ P_{\Omega} [g_{\Omega} ]^{n} \ ,
\label{S-n}
\end{equation}
 so that in the limit $n\rightarrow 1$ one recovers the Shannon entropy. That is, besides being an interesting information measure by itself, the R\'enyi-n entropy can also be useful to evaluate the Shannon entropy by means of a replica trick. For this purpose, one defines the replicated partition function

\begin{equation}
Z(n,\Omega )\ \equiv \ \int [Dg_{\Omega} ]\ ( P_{\Omega}[ g_{\Omega}])^{n} \ ,
\end{equation}then obtains
\begin{equation}
S( \Omega ) =\lim _{n\rightarrow 1} S^{( n)}( \Omega ) \ =\ -\left. \frac{\partial }{\partial n}\ln Z(n,\Omega )\right| _{n=1} \ .
\end{equation}
In other words, one first computes $\ln Z(n,\Omega )$ for integer $n$ and then analytically continues to $n\rightarrow 1$. In this context, the conditions for the uniqueness of the analytic continuation of a function defined only on the integers are discussed in Refs. \cite{Rubel1955,Cardy2008,Castro-Alvaredo_2008}. This reformulation turns the problem of computing the path integral of a logarithm into computing the logarithm of $\int P^{n}$, which, for Gaussian distributions, can be done by standard methods.

For general $n$, the R\'enyi-n entropy can be trivially obtained from the replicated partition function as follows
\begin{align}
    S^{(n)}(\Omega)=\frac{1}{1-n}\ln Z(n,\Omega)\;.\label{trivial-formula}
\end{align}
The replicated partition function for the vacuum state, denoted as $Z_{0}( n,\Omega )$, reads
\begin{align}
Z_{0}( n,\Omega ) &= \int [\mathcal{D} g_{\Omega} ]\ ( P_{0\Omega } [g_{\Omega} ])^{n} 
\nonumber \\
&= \left[\frac{\det (\mu ^{-1}\mathcal{O} )}{\det_{\overline{\Omega }} 
(\mu ^{-1}\mathcal{O} )}\right]^{\frac{n}{2}}\int [\mathcal{D} g_{\Omega} ] \ 
\exp\left( -n\int d^{d}\mathbf{x} \ \phi _{0\Omega} (\mathbf{x} )[\mathcal{O} \phi _{0\Omega} ](\mathbf{x} )\right) \;.
\end{align}
This Gaussian integral can be evaluated by shifting $O\rightarrow nO$ in $ P_{0\Omega } [g_{\Omega} ]$ and using the normalization condition $ \int [\mathcal{D} g_{\Omega} ]\ P_{0\Omega } [g_{\Omega} ]=1$. Then, $Z(n,\Omega )$ can be written in terms of the following ratios of determinants 
\begin{equation}\label{replicated partition function for vacuum state over Omega}
Z_{0}( n,\Omega ) =\left[\frac{\det\left( \mu ^{-1}\mathcal{O}\right)}{\det_{\overline{\Omega}}\left( \mu ^{-1}\mathcal{O}\right)}\right]^{n/2}\left[\frac{\det_{\overline{\Omega}}\left( \mu ^{-1} n\mathcal{O}\right)}{\det\left( \mu ^{-1} n\mathcal{O}\right)}\right]^{1/2} .
\end{equation}
Finally, taking $-\partial _{n}\ln Z(n,\Omega )$ at $n=1$ yields the Shannon entropy $ S_{0}( \Omega )$. Also, if we make the change $ \Omega \rightleftarrows \overline{\Omega }$, we have similar expressions for the R\'enyi-n entropy, $ S_{0}^{(n)}(\overline{\Omega })$, over the region $ \overline{\Omega }$. In a similar manner, to compute the R\'enyi-n entropy associated to the region $ \Omega \cup \overline{\Omega }$, we can use the replicated function, considering the probability density $\displaystyle P_{0}[ \phi ] =|\Psi _{0}[ \phi ] |^{2}$ calculated from the ground state wave functional \eqref{ground state wave functional}. Thus
\begin{equation}\label{replicated partition function for vacuum state over Omega and Omega bar}
Z_{0}( n,\Omega \cup \overline{\Omega }) =\int [\mathcal{D} \phi ]\ ( P_{0} [\phi ])^{n} =\frac{\left[\det\left( \mu ^{-1}\mathcal{O}\right)\right]^{n/2}}{\left[\det\left( n\mu ^{-1}\mathcal{O}\right)\right]^{1/2}}\;,
\end{equation}
which reproduces the standard result that in a free theory, the Shannon entropy is related to the spectrum of the operator $\mathcal{O}$.

To compute the R\'enyi-n entropy associated to the region $\Omega$ for the localized one-particle state, we consider the reduced probability density \eqref{reduced probability for the localized excited state}, so that the replicated partition function will be 
\begin{align}
\nonumber Z_{1} (n,\Omega )= & \int [\mathcal{D} g_{\Omega} ]( P_{1\Omega } [g_{\Omega} ])^{n}\\\label{Replicated partition function for the excited state}
= & 2^{n}\left[\frac{\det\left( \mu ^{-1}\mathcal{O}\right)}{\det_{\overline{\Omega}}\left( \mu ^{-1}\mathcal{O}\right)}\right]^{\frac{n}{2}}\int [\mathcal{D} g_{\Omega} ]\left[\left. \left(\frac{\delta ^{2}}{\delta \alpha ^{2}} -\frac{\delta ^{2}}{\delta \beta ^{2}}\right)\exp( W[\phi _{1\Omega} ,f_{\alpha ,\beta } ])\right| _{\alpha =\beta =0}\right]^{n} .
\end{align}
To calculate the integral $ \int [\mathcal{D} g_{\Omega} ]$, first, we must compute the variations of the exponential with respect to $ \alpha $ and $ \beta $ and evaluate these parameters at zero. Then, we expand the binomial expression. Now, introducing a generating functional, we can consider the even variations of this generating functional to compute the integral $ \int [\mathcal{D} g_{\Omega} ]$. This process enables us to obtain the following:

\begin{equation}\label{replicated partition function for localized excited state}
Z_{1} (n,\Omega )=Z_{0} (n,\Omega )F_{\Omega }^{( n)}( t) ,
\end{equation} where the factor correction $ F_{\Omega }^{( n)}( t)$ is a rather complex function involving integrals over $ \Omega $ of expressions with the wave packet function $ f( t,\mathbf{x})$ and operator inversions (for more details, see the Appendix \ref{Replicated Partition Function Calculation} or \cite{unesp2025_shannon}). This function encapsulates the excited-state contribution. Using the relation \eqref{trivial-formula} we obtain the following result for the R\'enyi-n entropy $S_1^{(n)}(\Omega)$
\begin{align}
    S_1^{(n)}(\Omega)= S_0^{(n)}(\Omega)+\Delta S^{(n)}(t,\Omega)\;,
\end{align}
where 
\begin{align}
    \Delta S^{(n)}(t,\Omega)=\frac{1}{1-n}\ln F_\Omega^{(n)}(t)
\end{align}
is the additional contribution to the entropy due to the one-particle excitation.
Again, taking the $n\rightarrow 1$ derivative of $\ln Z_{1} (n,\Omega )$ yields the Shannon entropy of the excited state. 

\begin{equation}\label{Shannon entropy for the localized excited state over Omega}
S_{1}( \Omega ) =S_{0}( \Omega ) +\Delta S( t,\Omega ) ,
\end{equation} where $ \Delta S( t,\Omega ) =-\left. \frac{\partial }{\partial n} F_{\Omega }^{( n)}( t)\right| _{n=1}$. Similar expressions are found for the entropies associated to the regions $\overline{\Omega}$ and $\Omega\cup\overline{\Omega}$. Physically, the particle excitation modifies the probability distribution for the field configurations, thereby altering the R\'enyi-$n$ entropies relative to the vacuum.

\subsection{R\'enyi-n Mutual Information Between Complementary Regions}

The R\'enyi-n mutual information 
\begin{equation}
I^{(n)}(\Omega:\overline{\Omega} )=S^{(n)}(\Omega)+S^{(n)}(\overline{\Omega} )-S^{(n)}(\Omega\cup \overline{\Omega} )
\end{equation}
is a measure of the correlation between the random variables corresponding to the values of the field in the two complementary spatial regions, $\Omega$ and $\overline{\Omega}$. Using Eqs. \eqref{trivial-formula}, \eqref{replicated partition function for localized excited state} we obtain the following result for the mutual information of the excited state
\begin{gather}
    I_1^{(n)}(\Omega,\bar{\Omega})=I_0^{(n)}(\Omega,\bar{\Omega})+\Delta I^{(n)}(t)\makebox[.5in]{,}\Delta I^{(n)}(t)=\frac{1}{1-n}\ln\left(\frac{F_\Omega^{(n)}F_{\overline{\Omega}}^{(n)}}{F^{(n)}_{\Omega\cup\overline{\Omega}}}\right)\;.
\end{gather}
Here, $I_0^{(n)}(\Omega,\bar{\Omega})$ is the R\'enyi-n mutual information for the vacuum state, given by
\begin{gather}
    I_0^{(n)}(\Omega,\bar{\Omega})=\frac{1}{1-n}\left(\frac{n}{2}\ln \frac{\det(\mu^{-1}\mathcal{O})}{\det_\Omega(\mu^{-1}\mathcal{O})\det_{\overline{\Omega}}(\mu^{-1}\mathcal{O})}-\frac{1}{2}\ln \frac{\det(\mu^{-1}n\mathcal{O})}{\det_\Omega(\mu^{-1}n\mathcal{O})\det_{\overline{\Omega}}(\mu^{-1}n\mathcal{O})}\right)\;. \label{renyi-vacuum}
\end{gather}
 In particular, we note that, using the results for the replicated partition functions of the vacuum state \eqref{replicated partition function for vacuum state over Omega} and \eqref{replicated partition function for vacuum state over Omega and Omega bar}, we find, for the vacuum Shannon mutual information 
\begin{equation}
I_{0} (\Omega,\overline{\Omega} )=\frac{1}{2}\ln\left(\frac{\det_{\Omega} (\mu ^{-1}\mathcal{O} )\det_{\overline{\Omega}} (\mu ^{-1}\mathcal{O} )}{\det (\mu ^{-1}\mathcal{O} )}\right) -\frac{1}{2}\left.\frac{\partial}{\partial n} \ln\left(\frac{\det_{\Omega} (n\mu ^{-1}\mathcal{O} )\det_{\overline{\Omega}} (n\mu ^{-1}\mathcal{O} )}{\det (n\mu ^{-1}\mathcal{O} )}\right)\right| _{n=1}\;,
\end{equation}
in agreement with the result obtained in Ref. \cite{David}. 

Furthermore, $\Delta I^{(n)}(t)$ denotes the additional contribution due to the particle, defined explicitly in Eq.~\eqref{corr-factor} of Appendix~\ref{Replicated Partition Function Calculation}. For a localized one-particle state, the R\'enyi-$n$ mutual information contains, in addition to the vacuum contribution, an extra term that depends on the wave-packet profile. In this work, we are primarily interested in these particle-induced contributions. To obtain quantitative results for them, we shall focus on the R\'enyi-2 (collision) entropy in the next section.

\section{Rényi-2 Entropy For the Localized One-particle State}
\label{sec:explicit}
To gain a quantitative understanding from the formal expression for the R\'enyi mutual information, we focus on a scenario that is tractable yet physically relevant. We will consider the R\'enyi-2 (collision) mutual information for a one-dimensional spatial setting where the complementary regions $\Omega$ and $\overline{\Omega}$ are defined as half-lines, $\Omega = ( -\infty ,0]$ and $\overline{\Omega} =[0,\infty) $, respectively. This simplified geometry makes the functional integrals and operator inversions manageable. Moreover, we will use a massless wave packet that minimizes the position velocity uncertainty $\displaystyle \Delta x\Delta v\geq \frac{1}{2} \langle \partial _{k} \omega _{k} \rangle $, which satisfies \eqref{Shrödiger equation to wavepacket} with $m=0$ and is given by (see \cite{localizedstates,ALHASHIMI20092599})
\begin{equation}\label{Lorentzian wave packet for tneq0}
f(t,x)=2i\sqrt{\frac{\alpha }{2\pi }}\frac{t-i\alpha }{(x-\langle x\rangle )^{2} -(t-i\alpha )^{2}} .
\end{equation}
We can see that at $\displaystyle t=0$, this wave packet is the Lorentzian wave packet
\begin{equation}
f(0,x)=\sqrt{\frac{\alpha }{2\pi }}\frac{2\alpha }{(x-\langle x\rangle )^{2} +\alpha ^{2}} ,
\end{equation}
 where $\alpha $ controls the width, and $\langle x\rangle $ is the average position of the particle. This setup allows for an explicit numerical investigation of how the particle's presence, specifically its position $\langle x\rangle $ and spatial width $\alpha $, modifies the mutual information between the two regions compared to that of the vacuum. 
\begin{figure}[ht]
    \centering
    
    \begin{subfigure}{0.48\textwidth}
        \centering
        \includegraphics[scale=0.4]{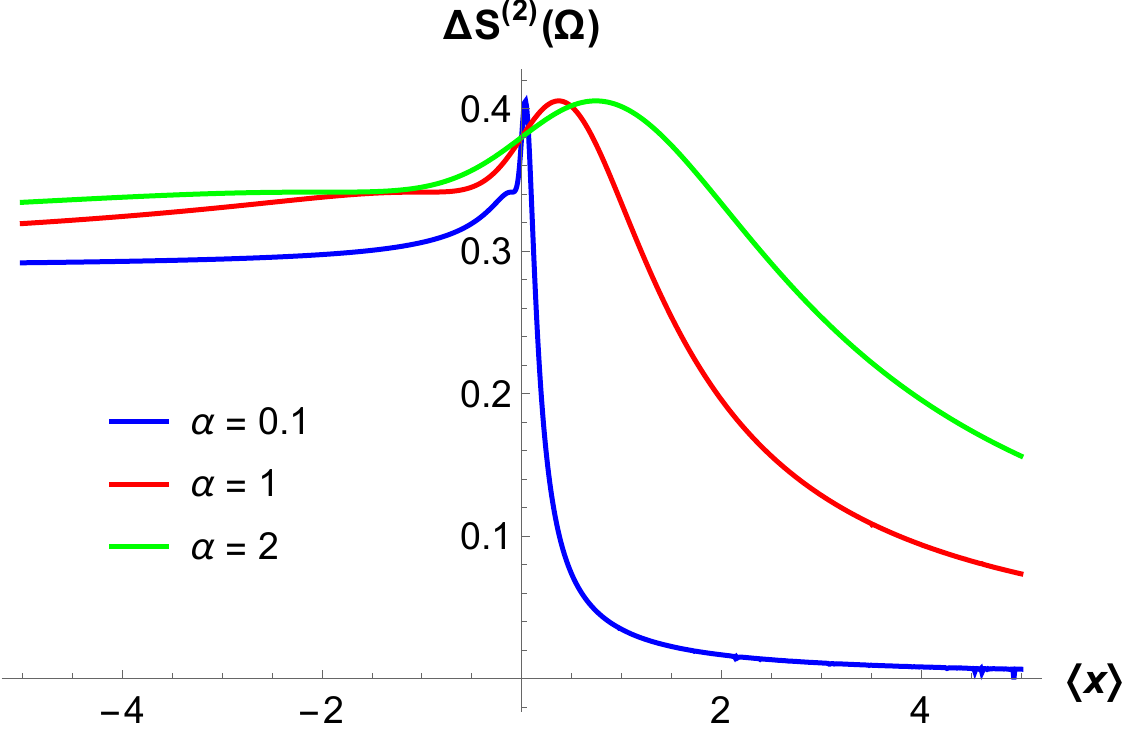}
        \caption{$\Omega$ region.}
        \label{fig:renyi-omega}
    \end{subfigure}
    \hfill
    \begin{subfigure}{0.48\textwidth}
        \centering
        \includegraphics[scale=0.4]{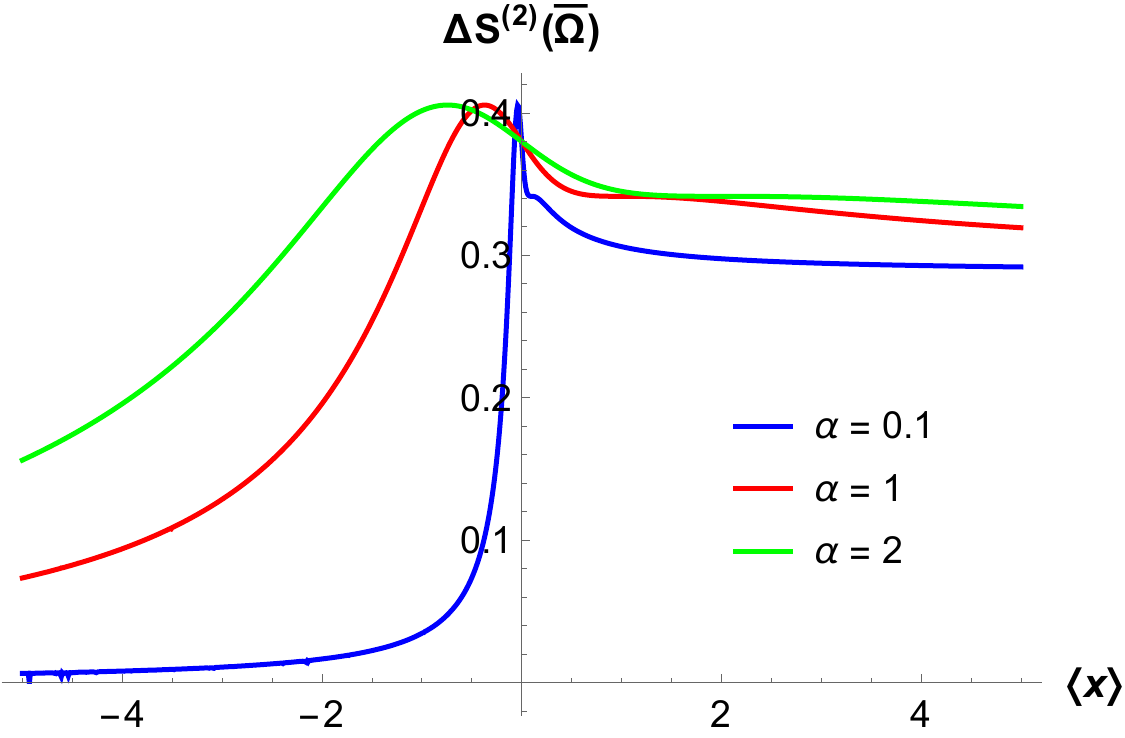}
        \caption{$\overline{\Omega}$ region.}
        \label{fig:renyi-omega-bar}
    \end{subfigure}

    \caption{Rényi-2 entropy corrections for the regions $\Omega$ and $\overline{\Omega}$ at $t=0$.}
    \label{fig:renyi-corrections}
\end{figure}

\subsection{Rényi entropy for the regions $\Omega=( -\infty ,0]$ and $\overline{\Omega} =[0,\infty) $}

Let us begin by discussing our results for the vacuum. In this case, it is convenient to first consider a massless scalar field defined on a line of length $L$, with $\Omega$ taken to be a subinterval of length $l$. The physical configuration of interest can then be obtained by setting $L = 2l$ and taking the limit $l \to \infty$. For this setup, we evaluate Eq.~\eqref{renyi-vacuum} with $n = 2$ for our chosen regions.
 To evaluate the functional determinant of $\mathcal{O}$ defined on an interval of length $l$ with Dirichlet boundary conditions on its endpoints, we may follow Ref. \cite{David} and use the $\zeta$ function regularization. In this setting, we define the spectral $\zeta_{A}$ function of a given differential operator $A$ as follows
\begin{gather}
\zeta_{A}(s)=\sum_i \lambda_i^{-s}\;, 
\end{gather}
where $\lambda_i$ are the eigenvalues of the operator. This defines an analytic function of $s$ for sufficiently large $s$. This function is then analytically continued to the vicinity of $s=0$, and one defines the logarithm of the functional determinant as
\begin{align}
    \ln \det A = -\zeta_{A}'(0)\;.
\end{align}
In the present case, the operator in question is a power of the Laplacian defined on a finite interval with length $l$, with the eigenfunctions satisfying Dirichlet boundary conditions at the endpoints. Then, the eigenvalues are simply
\begin{align}
    \lambda_n= \frac{n\pi}{\mu l}\;,
\end{align}
and the $\zeta$ function corresponding to the $\Omega$ region is
\begin{align}
    \zeta_{\Omega}(s)=\sum_n \left(\frac{n \pi}{\mu l}\right)^{-s}=\left(\frac{\pi}{\mu l}\right)^{-s}\zeta(s)\;, \label{ourzeta}
\end{align}
where $\zeta(s)$ is the usual Riemann $\zeta$ function. The sum over eigenvalues is well-defined for $Re(s)>1$, which is the original domain of definition of our spectral function $\zeta_\Omega(s)$. However, we can use \eqref{ourzeta}, together with the well-known analytic continuation of the Riemann $\zeta$ function, to analytically continue our spectral function to the whole complex plane, except for a simple pole at $s=1$. This allows us to evaluate the functional determinant
\begin{align}
    \ln \det_\Omega(\mu^{-1}\mathcal{O})=-\zeta_\Omega'(0)=\ln\left(\frac{\pi}{\mu l}\right)\zeta(0)-\zeta'(0)=\frac{1}{2} \ln (2\mu l)\;.
\end{align}
Similarly, the results for the regions $\overline{\Omega}$ and $\Omega \cup \overline{\Omega}$ are, respectively,
\begin{align}
    \ln \det_{\overline{\Omega}}(\mu^{-1}\mathcal{O})=\frac{1}{2} \ln(2\mu(L-l))\makebox[.5in]{,}\ln \det (\mu^{-1}\mathcal{O})=\frac{1}{2}\ln (2\mu L)\;.
\end{align}
The determinants for the operator $2\mathcal{O}$ can be easily obtained upon replacing $\mu \to \mu/2$ in the above expressions. Using these results, we are able to evaluate the R\'enyi-2 entropies for the vacuum, thus obtaining
\begin{gather}
    S_0^{(2)}(\Omega) = \frac{1}{8}\ln \frac{L-l}{L}\makebox[.5in]{,}S_0^{(2)}(\overline{\Omega}) = \frac{1}{8}\ln \frac{l}{L}\makebox[.5in]{,}S_0^{(2)}(\Omega\cup \overline{\Omega})=\frac{1}{8} \ln \left(\frac{1}{4\mu L}\right)\;.
\end{gather}
For the particular case where $\Omega$ is the half-line $[-l,0]$, we have $L=2l$, and therefore
\begin{gather}
    S_0^{(2)}(\Omega) = S_0^{(2)}(\overline{\Omega})=\frac{1}{8}\ln \frac{1}{2}\makebox[.5in]{,}S_0^{(2)}(\Omega\cup \overline{\Omega})=\frac{1}{8} \ln \left(\frac{1}{8\mu l}\right)\;. \label{renyi-2-vac}
\end{gather}
Let us now analyze the contributions due to the particle in the limit $l\to \infty$, initially considering $t=0$. The numerical evaluation of the correction to the Rényi-2 entropy, $\Delta S^{(2)}$, for each region as a function of the particle's initial position $\langle x\rangle $ reveals key behaviors, as illustrated in Figure \ref{fig:renyi-corrections}. When the particle is located well within a region (e.g., $\langle x\rangle \ll 0$ for region $\Omega$), the entropy correction $\Delta S^{(2)} (\Omega)$ is approximately constant and positive. A positive $\Delta S^{(2)}$ means the excitation introduces additional fluctuations in the likely field values within $\Omega$, effectively spreading out the reduced probability distribution. The magnitude of this effect depends on the wave packet's width, $\alpha $. Narrower wave packets (smaller $\alpha $) tend to produce smaller entropy corrections due to their more localized effect on the field. As the particle's average position approaches the boundary at $x=0$, the correction $\Delta S^{(2)}$ reaches a maximum, such that the broadest reduced probability distribution is achieved in this scenario. Conversely, when the particle moves far into the complementary region (e.g., $\langle x\rangle \gg 0$ for region $\Omega$), the correction $\Delta S^{(2)} (\Omega)$ decreases to zero. This is because the excitation's influence on the distant region becomes negligible, and the field configurations there are essentially those of the vacuum state.

 \begin{figure}[ht]
    \centering
     \includegraphics[scale=0.75]{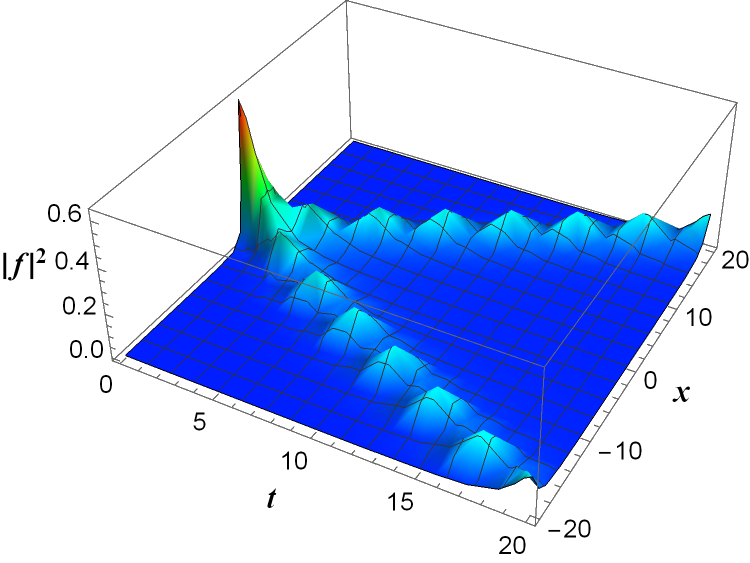}
\caption{Evolution of the squared absolute value of the wave packet $f(t,x)$ given by \eqref{Lorentzian wave packet for tneq0}, with $\alpha=1$ and $\langle x \rangle=0$.}
\label{Dynimical_Renyi2_entropy}
\end{figure}
\begin{figure}[ht]
    \centering
     \includegraphics[scale=0.25]{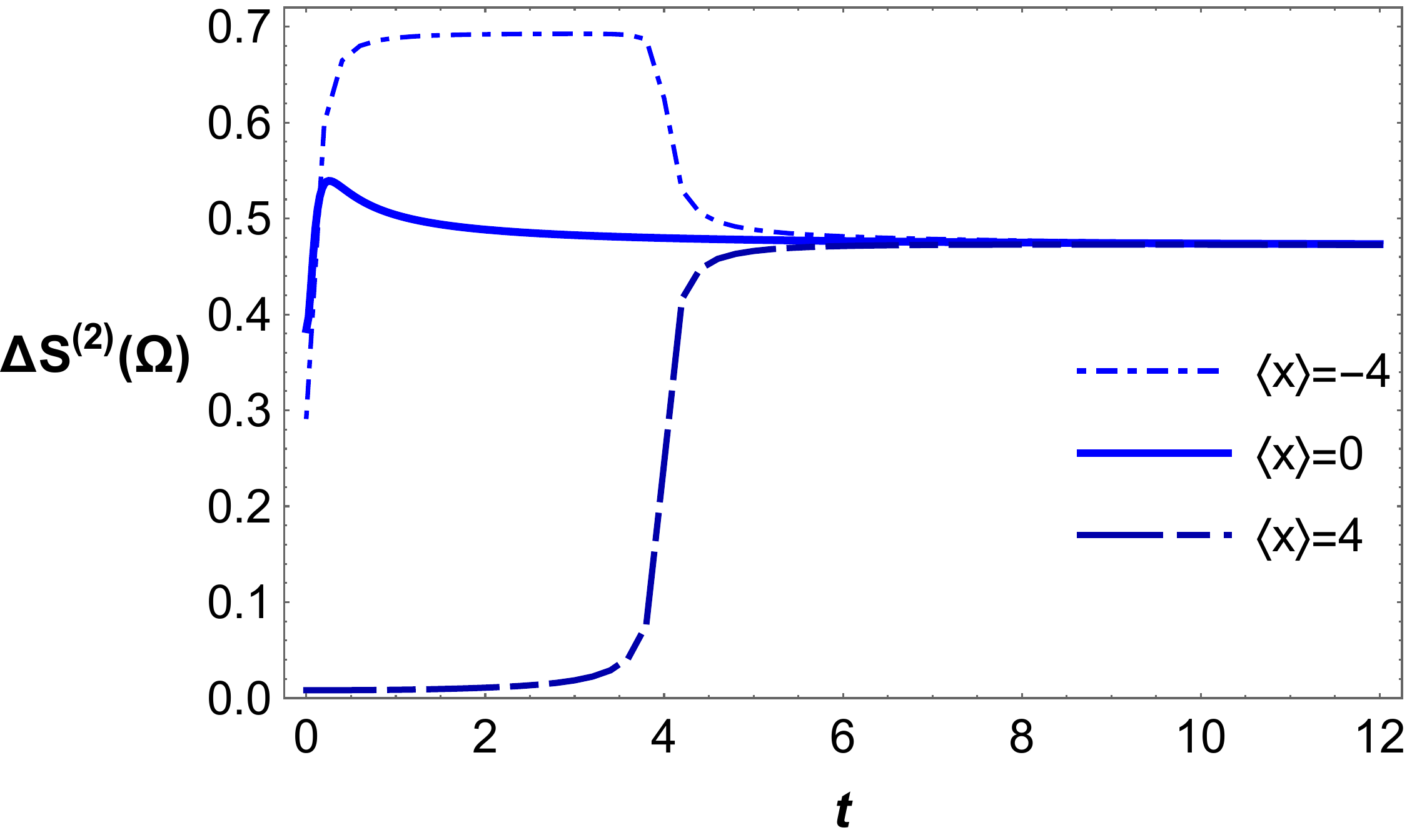}
\caption{Time evolution of the R\'enyi-2 entropy correction for the region $\Omega$ (with $\alpha=0.1$), for different values of $\langle x \rangle$.}
\label{Dynamical_Renyi2_entropy}
\end{figure}

As time passes, as shown in Figure \ref{Dynimical_Renyi2_entropy}, the wave packet progressively splits into two equally probable components moving in opposite directions. This splitting arises because the massless scalar field in 1+1 dimensions supports left- and right-moving modes that propagate without dispersion, causing an initially localized wave packet to evolve into two components traveling in opposite directions at the speed of light. The impact of this behavior on the Rényi-2 entropy is illustrated in Figure \ref{Dynamical_Renyi2_entropy}. When the particle originates within the region where entropy is measured, the additional R\'enyi-2 entropy is maximized immediately after the separation of the wave packet. As one part of the wave packet moves deeper into the region and the other crosses the boundary, the correction to the Rényi-2 entropy gradually decreases, eventually settling at a constant value determined solely by the inward-propagating portion. Furthermore, in the infinite-time limit, it can be verified that this constant value is independent of the wave packet width and is given by $\ln(8/5)$. If the particle begins at the boundary of the complementary regions, the correction reaches its maximum almost immediately and then decays as both components move away. Finally, when the particle originates outside the region, the initial correction at $t = 0$ is smaller and increases toward its asymptotic value as the wave packet enters the region.

\subsection{Rényi mutual information between the regions $\Omega$ and $\overline{\Omega}$}

\begin{figure}[ht]
    \centering
     \includegraphics[scale=0.4]{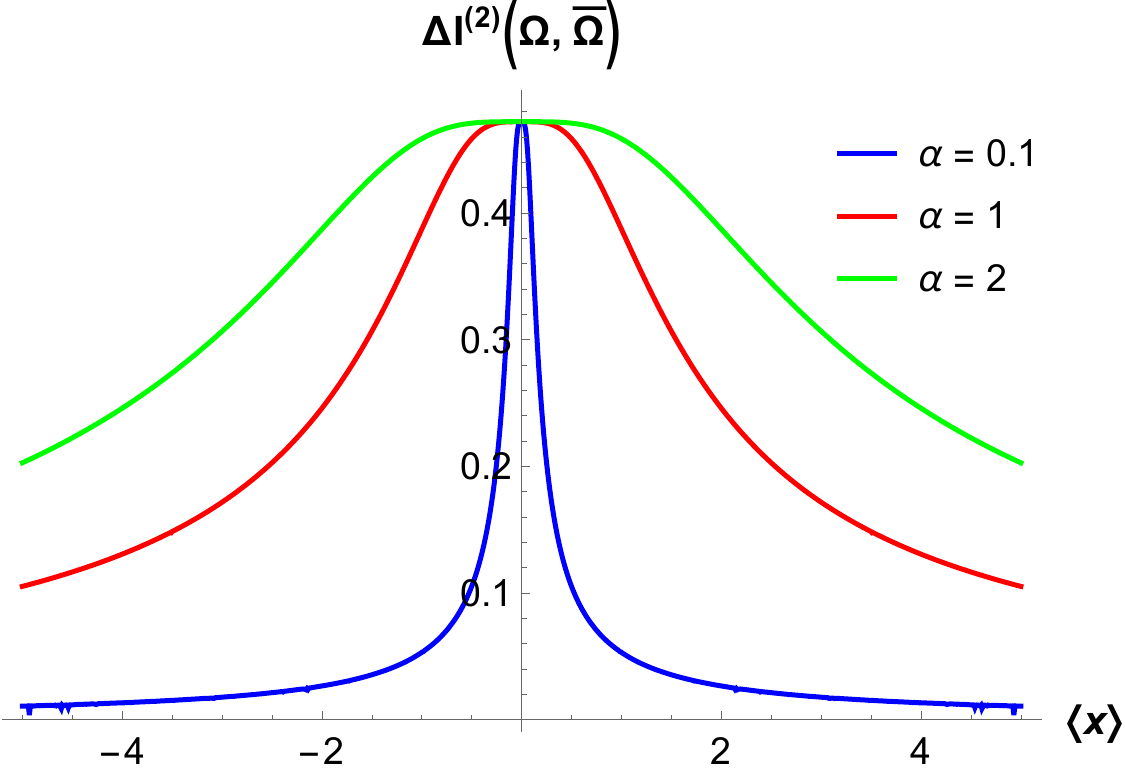}
\caption{The additional R\'enyi-2 mutual information due to the particle, at $t=0$.}
\label{Renyi mutual information between the negative and positive halfline}
\end{figure}

Let us initially note that we may obtain the mutual information of the vacuum immediately from Eq. \eqref{renyi-2-vac}. The result is
\begin{align}
    &I^{(2)}_0(\Omega,\bar{\Omega})=\frac{1}{8}\ln \left(\mu l\right)+\frac{1}{8}\ln 2\;.
\end{align}
In particular, we note that the vacuum mutual information diverges logarithmically with the system size, as expected.

Let us now examine the particle contributions, which are finite, independent of the regularization scheme, and therefore physically meaningful. The same was observed for the basis-independent entanglement measures studied in Refs. \cite{PhysRevLett.106.201601,palmai:2014,castroalvaredoI_2018,castroalvaredoII_2019,castroalvaredoIII_2019,Zhang2020, Zhang2021,Blanco2013,Wong2013} for a wide class of excited states. The numerical results for the additional Rényi-2 mutual information at $t=0$, shown in Figure \ref{Renyi mutual information between the negative and positive halfline}, demonstrate that the field correlations in the excited state, $I_{1}^{(2)}$, exceed those of the vacuum, $I_{0}^{(2)}$, by a finite amount. This excess information, $\Delta I^{(2)}$, quantifies the correlations introduced by the excitation, which are maximized when the particle is located precisely at the boundary ($\langle x\rangle =0$). At this point, the particle's wave packet is symmetrically distributed, maximizing the correlations between the field values in the two regions. The excess information decays rapidly with $\langle x \rangle$. The rate of this decay depends on the wave packet's width, $\alpha $: narrower wave packets lead to sharper decays.

\begin{figure}[ht]
    \centering
     \includegraphics[scale=0.7]{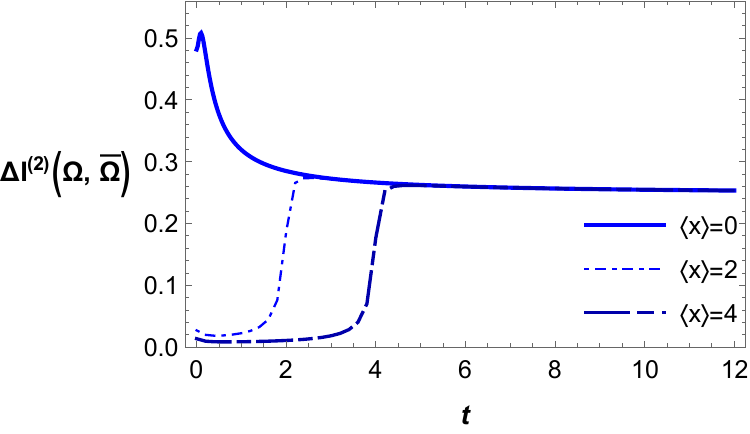}
\caption{The additional mutual information due to the particle as a function of time, with $\alpha=0.1$.}
\label{dyn-renyi2}
\end{figure}

Finally, we will address the evolution of the additional mutual information due to the particle over time, as illustrated in Figure \ref{dyn-renyi2}. When the particle begins within one of the complementary regions, the correction is generally initially smaller since its influence on the other region is limited. Once part of the wave packet crosses the boundary, however, the correction increases and reaches a maximum, in accordance with the intuitive picture that the correlations are maximized when the particle is localized near the boundary. As the two components move away from the boundary, the correction decreases and eventually approaches the constant asymptotic value $\ln(32/25)$. In this limit, one portion of the wave packet lies entirely within the region $\Omega$ and has no influence on $\bar{\Omega}$, while the complementary portion lies in $\bar{\Omega}$ and does not influence $\Omega$ at all. Nevertheless, the correlation between the field variables in $\Omega$ and $\bar{\Omega}$ remains nonzero. This can only occur if the local manner in which each excitation affects its own region is itself correlated with the other, despite their spatial separation. Since both excitations originate from the same initial wave packet at $t=0$, the resulting nonlocal correlation is strongly reminiscent of entanglement, albeit arising from a single-particle wave packet whose separated parts behave as if they were entangled with each other.
 It would be interesting to explore this further in future work, particularly by considering two-particle states in the Schr\"odinger representation.

\section{Conclusions}
\label{sec:conclusions}

We have analyzed the Rényi mutual information between complementary spatial regions in the presence of localized single-particle excitations of a free massless scalar field. Our results show that the mutual information consists of two distinct contributions: the geometric correlations already present in the vacuum, and a finite correction induced by the excitation, which depends on the wave-packet properties. This establishes that wave-packet excitations generate nontrivial spatial correlations across complementary regions, even in a free theory.

To gain a more quantitative assessment of the results, we focused on the Rényi-2 mutual information in one spatial dimension, where the regions are defined as half-lines. In this setting, we find that the particle-induced correlations are finite and positive. This is in sharp contrast with the vacuum correlations, which are divergent and regularization-dependent. That is, as observed for entanglement measures in excited states \cite{PhysRevLett.106.201601,palmai:2014,castroalvaredoI_2018,castroalvaredoII_2019,castroalvaredoIII_2019,Zhang2020,Zhang2021,Blanco2013,Wong2013}, particle-induced correlations encode relevant physical information. In addition, they are enhanced when the particle resides near the boundary. When the wave packet is localized deeper in the bulk, the corresponding correction decays, and the decay scale is set by the packet’s width, with narrower packets yielding correlations that are correspondingly more localized. We also examined the time evolution of these correlations. Due to the linear dispersion relation of the massless scalar field, an initially localized wave packet naturally splits into left- and right-moving components that propagate in opposite directions. Then, the mutual information reaches a maximum when there is maximum overlap of the excitation with the boundary, and decreases as the components separate. Interestingly, the particle-induced mutual information settles to a nonzero constant at late times, regardless of the initial position of the wave packet. Such behavior is possible only if the manner in which each excitation influences the surrounding field degrees of freedom remains correlated with that of the other, despite their spatial separation.
Since both excitations originate from the same initial wave packet at $t=0$, the resulting long-range correlation closely resembles entanglement, yet it emerges from a single-particle wave packet whose separated portions act as if they were entangled with one another.

In future work, we plan to extend this analysis to two-particle states, with the aim of developing a broader framework for understanding entanglement and mutual information in excited states formed by multiple wave packets in QFT.

\section{Acknowledgments}
This work was partially supported by Funda\c{c}\~ao de Amparo \`a Pesquisa do Estado de S\~ao Paulo (FAPESP), Grants No. 2023/18483-0 (D. R. J.), No. 2024/20896-3 (D. R. J.), No. 2018/25225-9 (G. K.) and No. 2021/14335
0 (G. K.); Conselho Nacional de Desenvolvimento Cient\'ifico e Tecnol\'ogico (CNPq) Grant No. 313254/2025-7 (G. K.); and Coordenac\c{c}\~ao de Aperfei\c{c}oamento de Pessoal de N{\'i}vel Superior (CAPES) Grants No. 88887.830725/2023-00 (W. A. I.) and 88887.183025/2025-00 (W. A. I.).

\appendix

\section{Replicated Partition Function Calculation}

\label{Replicated Partition Function Calculation}

This appendix breaks down the calculation of the replicated partition functions for a localized one-particle state. We'll cover the calculations for a subregion \(\Omega\), its complement \(\overline{\Omega}\), and the entire space \(\Omega \cup \overline{\Omega}\). These results are the basis for the Rényi entropies and mutual information discussed in the main body of the paper.

\subsection*{Replicated Partition Function for a Subregion \(\Omega\)}

The replicated partition function for an excited state in a subregion \(\Omega\) is given by:
\begin{equation}
Z_{1}(n,\Omega) = 2^{n}\left[\frac{\det\left( \mu^{-1}\mathcal{O}\right)}{\det_{\overline{\Omega}}\left( \mu^{-1}\mathcal{O}\right)}\right]^{\frac{n}{2}}\int [\mathcal{D} g_{\Omega} ]\left[\left. \left(\frac{\delta^{2}}{\delta \alpha^{2}} - \frac{\delta^{2}}{\delta \beta^{2}}\right)\exp(W[\phi_{1\Omega}, f_{\alpha,\beta}])\right| _{\alpha=\beta=0}\right]^{n},
\end{equation}
where \(f_{\alpha,\beta}(t,\mathbf{x}) = \alpha \ \Re[ f(t,\mathbf{x})] + i\beta \ \Im[f(t,\mathbf{x})]\) is the source term and \(W\) is the action:
\begin{equation} 
W[\phi_{1\Omega}, f_{\alpha,\beta}] \equiv -\int d^{d}\mathbf{x} \left( \phi_{1\Omega}(\mathbf{x})[\mathcal{O} \phi_{1\Omega}](\mathbf{x}) - f_{\alpha,\beta}(t,\mathbf{x})\left[\mathcal{O}^{1/2} \phi_{1\Omega}\right](\mathbf{x})\right). 
\end{equation} 
Our first step is to solve the functional integral over the boundary field \(g_{\Omega}\). To do this, we rewrite the action \(W\) in terms of \(g_{\Omega}\). This involves the fractional Laplacian operator \(\mathcal{O} := \sqrt{-\nabla^{2}}\), defined as:
\begin{equation}
[\mathcal{O} u](\mathbf{x}) = \text{P.V.} \int d^{d}\mathbf{y} \, K(\mathbf{x}-\mathbf{y})[u(\mathbf{x})-u(\mathbf{y})],
\end{equation}
with the kernel \(K\) given by:
\begin{equation} 
K(\mathbf{x}-\mathbf{y}) := \frac{\Gamma\left(\frac{d+1}{2}\right)}{\pi^{\frac{d+1}{2}}} \frac{1}{|\mathbf{x}-\mathbf{y}|^{d+1}} .
\end{equation}
Using solutions to the non-local Poisson problem, we can express the action \(W\) as a quadratic form in \(g_{\Omega}\):
\begin{equation}
W[g_{\Omega}, f_{\alpha,\beta}] = -\frac{1}{2}\int_{\Omega} d^{d}\mathbf{x} \, g_{\Omega}(\mathbf{x})\int_{\Omega} d^{d}\mathbf{z} \, g_{\Omega}(\mathbf{z}) \, \hat{o}(\mathbf{x},\mathbf{z}) + \int_{\Omega} d^{d}\mathbf{z} \, g_{\Omega}(\mathbf{z}) \, a_{f_{\alpha \beta}}(t,\mathbf{z}) + b_{f_{\alpha \beta}}(t),
\end{equation}
where we've defined the following shorthand operators and terms:
\begin{itemize}
    \item \textbf{The Kernel \(\hat{o}(\mathbf{x},\mathbf{z})\):}\begin{equation}
        \frac{1}{2}\hat{o}(\mathbf{x},\mathbf{z}) \equiv [\mathcal{O}_{\Omega}\delta](\mathbf{x}-\mathbf{z}) + \delta(\mathbf{x}-\mathbf{z})\int_{\overline{\Omega}} d^{d}\mathbf{y} \, K(\mathbf{x}-\mathbf{y}) - \int_{\overline{\Omega}} d^{d}\mathbf{y} \, K(\mathbf{x}-\mathbf{y})\mathcal{P}(\mathbf{z},\mathbf{y})
    \end{equation}
    
    \item \textbf{The Source Term \(a_f(\mathbf{z})\):} \begin{align}
       \nonumber a_{f}(t,\mathbf{z}) \equiv & \left[\mathcal{O}^{1/2} f\right](t,\mathbf{z}) + \frac{1}{2}\int_{\overline{\Omega}} d^{d}\mathbf{y} \, K(\mathbf{z}-\mathbf{y}) \int_{\overline{\Omega}} d^{d}\mathbf{x} \, G(\mathbf{y},\mathbf{x})\left[\mathcal{O}^{1/2} f\right](t,\mathbf{x})\\
        &+ \frac{1}{2}\int_{\overline{\Omega}} d^{d}\mathbf{x} \, \mathcal{P}(\mathbf{z},\mathbf{x})\left[\mathcal{O}^{1/2} f\right](t,\mathbf{x})
    \end{align}
   
    \item \textbf{The Constant Term \(b_f\):} \begin{equation}
        b_{f}(t) \equiv \frac{1}{4}\int_{\overline{\Omega}} d^{d}\mathbf{x} \int_{\overline{\Omega}} d^{d}\mathbf{y} \, G(\mathbf{x},\mathbf{y})\left[\mathcal{O}^{1/2} f\right](t,\mathbf{y})\left[\mathcal{O}^{1/2} f\right](t,\mathbf{x})
    \end{equation}
\end{itemize}
Here, \(\mathcal{O}_{\Omega}\) is the fractional Laplacian restricted to \(\Omega\), \(G(\mathbf{x},\mathbf{y})\) is the Green's function for the problem, and \(\mathcal{P}(\mathbf{z},\mathbf{y})\) is the Poisson kernel.

After differentiating with respect to \(\alpha\) and \(\beta\) and setting them to zero, the term in the partition function becomes:
\begin{equation} 
\exp(W[g_{\Omega}])\left[\left| \int_{\Omega} d^{d}\mathbf{z} \, g(\mathbf{z}) a_{f}(\mathbf{z})\right|^{2} + 2b_{R} + 2b_{I}\right], 
\end{equation}
where \(W[g_{\Omega}]\) is the quadratic part of the action, and \(b_R\) and \(b_I\) are derived from the real and imaginary parts of the wave packet \(f(t,\mathbf{x})\).

The replicated partition function integral is now:
\begin{align}
Z_{1}(n,\Omega) &= 2^{n}\left[\frac{\det\left(\mu^{-1}\mathcal{O}\right)}{\det_{\overline{\Omega}}\left(\mu^{-1}\mathcal{O}\right)}\right]^{\frac{n}{2}} \nonumber \\
&\times \int [\mathcal{D} g_{\Omega}] \, \exp(nW[g_{\Omega}])\left(\left| \int_{\Omega} d^{d}\mathbf{z} \, g(\mathbf{z}) a_{f}(\mathbf{z})\right|^{2} + 2b_{R} + 2b_{I}\right)^{n} .
\end{align}
To solve this Gaussian integral with a polynomial prefactor, we introduce a generating functional \(\Gamma_{\Omega}[J]\):
\begin{equation}
\Gamma_{\Omega}[J] \equiv \int [\mathcal{D} g_{\Omega}]\exp\left( -\frac{n}{2}\int_{\Omega} d^{d}\mathbf{x} \int_{\Omega} d^{d}\mathbf{z} \, g_{\Omega}(\mathbf{x}) g_{\Omega}(\mathbf{z})\hat{o}(\mathbf{x},\mathbf{z}) + \int_{\Omega} d^{d}\mathbf{x} \, g_{\Omega}(\mathbf{x}) J(\mathbf{x})\right). 
\end{equation}
This is a standard Gaussian functional integral with a known solution:
\begin{equation}
\Gamma_{\Omega}[J] = [\det(n\hat{o})]^{-1/2}\exp\left(\frac{1}{2n}\int_{\Omega}d^{d}\mathbf{x} \int_{\Omega}d^{d}\mathbf{y} \, J(\mathbf{x}) J(\mathbf{y}) \, \hat{o}^{-1}(\mathbf{x},\mathbf{y})\right).
\end{equation}
Any integral involving powers of \(g_{\Omega}\) can now be found by taking functional derivatives of \(\Gamma_{\Omega}[J]\) with respect to the source \(J(\mathbf{x})\) and then setting \(J=0\). The general result for an even number of derivatives, which is what we need, is given by a sum over pairings (similar to Wick's theorem):
\begin{equation}
\frac{\delta^{2m} \Gamma_{\Omega}[J]}{\delta J(\mathbf{x}_{1})\cdots \delta J(\mathbf{x}_{2m})}\Bigg|_{J=0} = \frac{[\det(n\hat{o})]^{-1/2}}{n^{m}} \sum_{\text{permutations}} \hat{o}_{s}^{-1}(\mathbf{x}_{k_{1}},\mathbf{x}_{k_{2}})\cdots \hat{o}_{s}^{-1}(\mathbf{x}_{k_{2m-1}},\mathbf{x}_{k_{2m}}) ,
\end{equation}
where \(\hat{o}_{s}^{-1}\) is the symmetric part of the inverse operator \(\hat{o}^{-1}\), and the sum has \((2m-1)!!\) terms.

By inserting this back into the expression for \(Z_1(n,\Omega)\), using the binomial theorem, we have
\begin{align} \label{eq:11_app}
Z_{1}(n,\Omega) = & \, 2^{n}\left[\frac{\det\left(\mu^{-1}\mathcal{O}\right)}{\det_{\overline{\Omega}}\left(\mu^{-1}\mathcal{O}\right)}\right]^{\frac{n}{2}} [\det(\hat{o})]^{-1/2} \sum_{m=1}^{n}\binom{n}{m} 2^{n-m}(b_{R} + b_{I})^{n-m}\frac{1}{n^{m}} \nonumber \\
& \times \sum_{\text{permut}}\int_{\Omega}d^{d}\mathbf{z}_{1}\cdots d^{d}\mathbf{z}_{2m} \, a_{f}(\mathbf{z}_{1})\cdots a_{f}(\mathbf{z}_{m}) a_{f^{*}}(\mathbf{z}_{m+1})\cdots a_{f^{*}}(\mathbf{z}_{2m}) \nonumber \\
& \times \hat{o}_{s}^{-1}(\mathbf{z}_{k_{1}},\mathbf{z}_{k_{2}})\cdots \hat{o}_{s}^{-1}(\mathbf{z}_{k_{2m-1}},\mathbf{z}_{k_{2m}})
\end{align}
We can note that
\begin{align} \label{eq:12_app}
& \sum_{\text{permut}}\int_{\Omega}d^{d}\mathbf{z}_{1}\cdots d^{d}\mathbf{z}_{2m} \, a_{f}(\mathbf{z}_{1})\cdots a_{f}(\mathbf{z}_{m}) a_{f^{*}}(\mathbf{z}_{m+1})\cdots a_{f^{*}}(\mathbf{z}_{2m}) \nonumber \\
& \times \hat{o}_{s}^{-1}(\mathbf{z}_{k_{1}},\mathbf{z}_{k_{2}})\cdots \hat{o}_{s}^{-1}(\mathbf{z}_{k_{2m-1}},\mathbf{z}_{k_{2m}}) \nonumber \\
& = \sum_{2r+l=m} C_{rl}\left| \int_{\Omega} d^{d}\mathbf{x} \, a_{f}(\mathbf{x})\left[\hat{o}_{s}^{-1} a_{f}\right](\mathbf{x})\right|^{2r} \left(\int_{\Omega} d^{d}\mathbf{x} \, a_{f}(\mathbf{x})\left[\hat{o}_{s}^{-1} a_{f^{*}}\right](\mathbf{x})\right)^{l}
\end{align}
where $C_{rl}$ are coefficients that satisfy
\begin{equation}
\sum_{2r+l=m} C_{rl} = (2m-1)!!\;.
\end{equation}
Then, inserting the general term \eqref{eq:12_app} in \eqref{eq:11_app}, we obtain
\begin{align} \label{eq:13_app}
Z_{1}(n,\Omega) = & \, 2^{n}\left[\frac{\det\left(\mu^{-1}\mathcal{O}\right)}{\det_{\overline{\Omega}}\left(\mu^{-1}\mathcal{O}\right)}\right]^{\frac{n}{2}} [\det(\hat{o})]^{-1/2} \sum_{m=1}^{n}\binom{n}{m} 2^{n-m}(b_{R} + b_{I})^{n-m}\frac{1}{n^{m}} \nonumber \\
& \times \sum_{2r+l=m} C_{rl}\left| \int_{\Omega} d^{d}\mathbf{x} \, a_{f}(\mathbf{x})\left[\hat{o}_{s}^{-1} a_{f}\right](\mathbf{x})\right|^{2r}\left(\int_{\Omega} d^{d}\mathbf{x} \, a_{f}(\mathbf{x})\left[\hat{o}_{s}^{-1} a_{f^{*}}\right](\mathbf{x})\right)^{l}.
\end{align}
Now, considering the normalization conditions
\begin{equation}
\int [\mathcal{D} g_{\Omega}] P_{0\Omega}[g_{\Omega}] = 1, \qquad \int [\mathcal{D} g_{\Omega}] P_{1\Omega}[g_{\Omega}] = 1,
\end{equation}
and
\begin{equation}
\det(\hat{o}) = \left[\frac{\det\left(\mu^{-1}n\mathcal{O}\right)}{\det_{\overline{\Omega}}\left(\mu^{-1}n\mathcal{O}\right)}\right], \qquad \int_{\Omega}d^{d}\mathbf{x} \, a_{f}(\mathbf{x})\left[\hat{o}_{s}^{-1} a_{f^{*}}\right](\mathbf{x}) = \frac{1}{2} - 2(b_{R} + b_{I}), 
\end{equation}
and using the binomial theorem again, we obtain for \eqref{eq:13_app}: 
\begin{equation}
Z_{1}(n,\Omega) = Z_{0}(n,\Omega) \, F_{\Omega}^{(n)}(t),
\end{equation}
where \(Z_{0}(n,\Omega)\) is the replicated partition function for the vacuum state, 
and \(F_{\Omega}^{(n)}(t)\) is the correction factor: 
\begin{align}
    F_{\Omega}^{(n)}(t) &= \sum_{m=0}^{n}\binom{n}{m}\sum_{2r+l=m} C_{rl}
    \left| \int_{\Omega} d^{d}\mathbf{x} \, a_{f}(\mathbf{x})
    \left[\hat{o}_{s}^{-1} a_{f}\right](\mathbf{x})\right|^{2r}\nonumber \\
    &\times \sum_{k=0}^{l}(-1)^{k}\binom{l}{k}(b_{R} + b_{I})^{n-m+k}\frac{2^{2n+k-m-l}}{n^{m}}. \label{corr-factor}
\end{align}
The combinatorial coefficients \(C_{rl}\) result from grouping the terms in the permutation sum. The calculation for the complementary region, \(Z_{1}(n,\overline{\Omega})\), follows by simply swapping \(\Omega \leftrightarrow \overline{\Omega}\).

\subsection*{Replicated Partition Function for the Full Space \(\Omega \cup \overline{\Omega}\)}

For the full space, the calculation is simpler as there are no boundaries. The partition function is given by:
\begin{equation}
Z_{1}(n, \Omega \cup \overline{\Omega}) = \int [\mathcal{D} \phi] \, (P_{1}[\phi])^{n} ,
\end{equation}
where \(P_1[\phi]\) is the probability distribution for the single-particle state \eqref{excited-state probability}. We again use a generating functional, this time for the field \(\phi\) over the entire space. Thus, we define
\begin{equation}
\Gamma_{\Omega \cup \overline{\Omega}}[J] \equiv \int [\mathcal{D}\phi] \exp\left(-n\int d^{d}\mathbf{x} \, \phi(\mathbf{x})[\mathcal{O}\phi](\mathbf{x}) + \int d^{d}\mathbf{x} \, \phi(\mathbf{x})J(\mathbf{x})\right).
\end{equation}
We can express the replicated partition function in terms of the variation of this generating functional. Therefore
\begin{align}
\nonumber Z_{1}(n,\Omega \cup \overline{\Omega}) = & \, 2^{n}\left[\det\left(\mu^{-1}\mathcal{O}\right)\right]^{\frac{n}{2}}\int d^{d}\mathbf{x}_{1}\cdots d^{d}\mathbf{x}_{2n} \left[\mathcal{O}^{\frac{1}{2}} f\right](t,\mathbf{x}_{1})\cdots \left[\mathcal{O}^{\frac{1}{2}} f\right](t,\mathbf{x}_{n}) \\\label{eq:15_app}
& \times \left[\mathcal{O}^{\frac{1}{2}} f^{*}\right](t,\mathbf{x}_{n+1})\cdots \left[\mathcal{O}^{\frac{1}{2}} f^{*}\right](t,\mathbf{x}_{2n}) \left. \frac{\delta^{2n} \Gamma_{\Omega \cup \overline{\Omega}}[J]}{\delta J(\mathbf{x}_{1})\cdots \delta J(\mathbf{x}_{2n})}\right|_{J=0}.
\end{align}
The $2m$-th variation of $\Gamma_{\Omega \cup \overline{\Omega}}[J]$ will be composed of products of $m$ disjoint operators $\mathcal{O}^{-1}$
\begin{align}
\frac{\delta^{2m} \Gamma_{\Omega \cup \overline{\Omega}}[J]}{\delta J(\mathbf{x}_{1})\cdots 
\delta J(\mathbf{x}_{2m})}\Bigg|_{J=0} &= \left[\det\left(\mu^{-1}n\mathcal{O}\right)\right]^{-1/2}
\nonumber \\
&\times \frac{1}{(2n)^{m}} \sum_{\text{permut}}\mathcal{O}^{-1}(\mathbf{x}_{k_{1}},\mathbf{x}_{k_{2}})
\cdots \mathcal{O}^{-1}(\mathbf{x}_{k_{2m-1}},\mathbf{x}_{k_{2m}}).
\end{align}
Thus, taking into account these variations in \eqref{eq:15_app} (with $m=n$), we have
\begin{align}
Z_{1}(n,\Omega \cup \overline{\Omega}) = & \, \frac{1}{n^{n}}\frac{\left[\det\left(\mu^{-1}\mathcal{O}\right)\right]^{n/2}}{\left[\det\left(\mu^{-1}n\mathcal{O}\right)\right]^{1/2}} \sum_{2r+l=n} C_{rl}\left| \int d^{d}\mathbf{x} \, \left[\mathcal{O}^{\frac{1}{2}} f\right](t,\mathbf{x}) \left[\mathcal{O}^{-\frac{1}{2}} f\right](t,\mathbf{x})\right|^{2r} \nonumber \\
& \times \left(\int d^{d}\mathbf{x} \, \left[\mathcal{O}^{\frac{1}{2}} f\right](t,\mathbf{x}) \left[\mathcal{O}^{-\frac{1}{2}} f^{*}\right](t,\mathbf{x})\right)^{l}.
\end{align}
Taking into account the normalization condition $\int d^{d}\mathbf{x} \, |f(t,\mathbf{x})|^{2} = 1$. This leads to the result \begin{equation}
    Z_{1}(n,\Omega \cup \overline{\Omega}) = Z_{0}(n,\Omega \cup \overline{\Omega}) F_{\Omega \cup \overline{\Omega}}^{(n)}(t),
\end{equation} where \(Z_{0}\) is the vacuum partition function, and the correction factor is \begin{equation}
    F_{\Omega \cup \overline{\Omega}}^{(n)}(t) = \frac{1}{n^{n}}\sum_{2r+l=n} C_{rl}\left| \int d^{d}\mathbf{x} \, f^{2}(t,\mathbf{x})\right|^{2r}\;.
\end{equation}

\subsection*{Explicit Results for \(n=2\)}

The general expressions for \(F^{(n)}\) are quite involved. However, for \(n=2\) (which is used to find the collision entropy), they simplify considerably. The correction factors become:
\begin{align}
F_{\Omega}^{(2)}(t) &= \left| \int_{\Omega} d^{d}\mathbf{x} \, a_{f}(t,\mathbf{x})\left[\hat{o}_{s}^{-1} a_{f}\right](t,\mathbf{x})\right|^{2} + \frac{1}{2} + 8(b_{R}(t) + b_{I}(t))^{2} \\
F_{\overline{\Omega}}^{(2)}(t) &= \left| \int_{\overline{\Omega}} d^{d}\mathbf{x} \, \overline{a}_{f}(t,\mathbf{x})\left[\overline{\hat{o}}_{s}^{-1}\overline{a}_{f}\right](t,\mathbf{x})\right|^{2} + \frac{1}{2} + 8(\overline{b}_{R}(t) + \overline{b}_{I}(t))^{2} \\
F_{\Omega \cup \overline{\Omega}}^{(2)}(t) &= \frac{1}{4}\left(\left| \int d^{d}\mathbf{x} \, f^{2}(t,\mathbf{x})\right|^{2} + 2\right)\;.
\end{align}
In these expressions, the barred functions (\(\overline{a}_{f}\), \(\overline{b}_{f}\), etc.) are defined for the complementary region \(\overline{\Omega}\). These simplified formulas are used for the numerical computation of the Renyi-2 mutual information in the main text.

\subsubsection*{One Dimension Case with $\Omega =(-\infty ,0]$ and $\overline{\Omega } =[ 0,\infty)$}

To obtain explicit results in the one-dimensional case, we consider the regions to be half-lines, specifically $\Omega =(-\infty ,0]$ and $\overline{\Omega } =[ 0,\infty )$. For this configuration, the necessary components of our calculations are given by the following expressions:

\begin{itemize}
    \item \textbf{The Poisson Kernel:} The Poisson kernels for the complementary regions are defined as:
    \begin{gather*}
        \mathcal{P}( z,x) =\frac{1}{\pi }\sqrt{\frac{x}{|z|}}\frac{1}{|x-z|} ,\quad z\leq 0,x\geq 0. \\
        \overline{\mathcal{P}}( z,y) =\frac{1}{\pi }\sqrt{\frac{|y|}{z}}\frac{1}{|x-z|} ,\quad z\geq 0,y\leq 0.
    \end{gather*}

    \item \textbf{The Green's Function:} The Green's functions for the complementary regions are given by:
    \begin{align}
        G( x,y) &=\begin{cases}
            \frac{2}{\pi }\operatorname{arctanh}\left(\sqrt{\frac{y}{x}}\right) , & x >y, \\
            \frac{2}{\pi }\operatorname{arctanh}\left(\sqrt{\frac{x}{y}}\right) , & x< y, 
        \end{cases} \\
        G( |x|,|y|) &=\overline{G}( x,y).
    \end{align}
\end{itemize}

Considering the wave packet given by equation \eqref{Lorentzian wave packet for tneq0}, we find that:
\begin{equation}
    \left[ O^{1/2} f\right]( t,x) =\frac{1}{2}\sqrt{\frac{\alpha }{2}}\left[\frac{1}{[ \alpha +i( x-\langle x\rangle +t)]^{3/2}} +\frac{1}{[ \alpha -i( x-\langle x\rangle -t)]^{3/2}}\right].
\end{equation}
Given this, we can evaluate the following terms:
\begin{itemize}
    \item \textbf{The Source Terms $\boldsymbol{a_{f}( z)}$ and $\boldsymbol{\overline{a}_{f}( z)}$:}
    \begin{align}
        a_{f}( t,z) &=\left[ O^{1/2} f\right] (t,z)+\int _{0}^{\infty } dx\ \left[ O^{1/2} f\right]( t,x) \ \mathcal{P} (z,x),\quad z< 0 ,\\
        \overline{a}_{f}( t,z) &=\left[ O^{1/2} f\right] (t,z)+\int _{-\infty }^{0} dx\ \left[ O^{1/2} f\right]( t,x) \ \overline{\mathcal{P}} (z,x),\quad z >0.
    \end{align}

    \item \textbf{The Constant Terms $\boldsymbol{b_{f}}$ and $\boldsymbol{\overline{b}_{f}}$:}
    \begin{align}
        b_{f}( t) &=\frac{1}{4}\int _{0}^{\infty } dx\int _{0}^{\infty } dy\ G( x,y) \ \left[ O^{1/2} f\right]( t,y) \ \left[ O^{1/2} f\right] (t,x), \\
        \overline{b}_{f}( t) &=\frac{1}{4}\int _{-\infty }^{0} dx\int _{-\infty }^{0} dy\ \overline{G}( x,y) \ \left[ O^{1/2} f\right]( t,y) \ \left[ O^{1/2} f\right] (t,x).
    \end{align}
\end{itemize}

Finally, the inverse operators $\hat{o}_{s}^{-1}$ and $\overline{\hat{o}}_{s}^{-1}$ are given by:
\begin{align}
    \left[\hat{o}_{s}^{-1} f\right]( t,x) &=-\frac{1}{2\pi }\left(\int _{-\infty }^{x} dz\ f( t,z)\ln\left( 1-\frac{x}{z}\right) +\int _{x}^{0} dz\ f( t,z)\ln\left(\frac{x}{z} -1\right)\right) ,\\
    \left[\overline{\hat{o}}_{s}^{-1} f\right]( t,x) &=-\frac{1}{2\pi }\left(\int _{x}^{\infty } dz\ f( t,z)\ln\left( 1-\frac{x}{z}\right) +\int _{0}^{x} dz\ f( t,z)\ln\left(\frac{x}{z} -1\right)\right).
\end{align}
Therefore, the correction factors for this specific one-dimensional case are:
\begin{align}
    F_{\Omega}^{(2)} (t)=&\left| \int _{-\infty }^{0} dx\ a_{f} (t,x)\left[\hat{o}_{s}^{-1} a_{f}\right] (t,x)\right| ^{2} +\frac{1}{2} +8( b_{R}( t) +b_{I}( t))^{2}, \\
    F_{\overline{\Omega }}^{(2)} (t)=&\left| \int _{0}^{\infty } dx\ \overline{a}_{f} (t,x)\left[\overline{\hat{o}}_{s}^{-1}\overline{a}_{f}\right] (t,x)\right| ^{2} +\frac{1}{2} +8(\overline{b}_{R}( t) +\overline{b}_{I}( t))^{2}, \\
    F_{\Omega \cup \overline{\Omega }}^{(2)} (t)=&\frac{1}{4}\left(\left| \int _{-\infty }^{\infty } dx\ f^{2}( t,x)\right| ^{2} +2\right)=\frac{1}{4}\left[\frac{\alpha ^{2}}{t^{2} +\alpha ^{2}} +2\right], 
\end{align}
where $b_R$ and $b_T$ are given by
\begin{align}
&b_{R} (t)=\frac{1}{\pi }\int _{0}^{\infty } dy\int _{0}^{y} dx\operatorname{arctanh}\left(\sqrt{\tfrac{x}{y}}\right)\left[ {\rm Re}(\mathcal{O}^{1/2}f)\right] (t,y)\left[{\rm Re}(\mathcal{O}^{1/2}f)\right] (t,x)\;,\\
& b_{I} (t)=\frac{1}{\pi }\int _{0}^{\infty } dy\int _{0}^{y} dx\operatorname{arctanh}\left(\sqrt{\tfrac{x}{y}}\right)\left[{\rm Im}(\mathcal{O}^{1/2}f)\right] (t,y)\left[{\rm Im}(\mathcal{O}^{1/2}f)\right] (t,x)\;.
\end{align}

Finally, we note that from the correction factor $F_{\Omega\cup \overline{\Omega}}^{(2)} (t)$, corresponding to the union of the two complementary regions, one observes that the correction to the Rényi-2 entropy, namely:
\begin{equation}
\Delta S^{(2)}(\Omega\cup\overline{\Omega})=-\ln\left[\frac{1}{4}\left(\frac{\alpha ^{2}}{t^{2} 
+\alpha ^{2}} +2\right)\right],
\end{equation}
grows monotonically with time. This behavior contrasts with that of the correction to the Rényi entropy restricted to the region $\Omega$, which instead decreases and asymptotically approaches a fixed value as time progresses. This feature is reminiscent of the second law of thermodynamics, according to which the entropy of an isolated system evolves irreversibly towards larger values.

\end{document}